# Key management system for WSNs based on hash functions and elliptic curve cryptography

Hamzeh Ghasemzadeh
Electrical Engineering Department
IAU, Damavand Branch
Tehran, Iran
hamzeh_g62@yahoo.com

Ali Payandeh
ICT Department
Malek-e-Ashtar University
Tehran, Iran
payandeh@mut.ac.ir

Mohammad Reza Aref
Electrical Engineering Department
Sharif University of technology
Tehran, Iran
aref@sharif.edu

*Abstract*— Due to hostile environment and wireless communication channel, security mechanisms are essential for wireless sensor networks (WSNs). Existence of a pair of shared key is a prerequisite for many of these security mechanisms; a task that key management system addresses. Recently, an energy efficient method based on public key cryptography (PKC) was proposed. We analyze this protocol and show that it is vulnerable to denial of service (DOS) attacks and adversary can exhaust memory and battery of nodes. Then, we analyze this protocol and show that using a more knowledgeable BS this vulnerability can be solved very efficiently. Based on this observation we propose a modified version of the protocol that achieves immediate authentication and can prevent DOS attacks. We show that the improved protocol achieves immediate authentication at the expense of 1.82 mj extra energy consumption while retaining other desirable characteristics of the basic method.[1]

I. INTRODUCTION

Wireless sensor Network (WSN) is a collection of large number of low-power sensor actuator nodes and its main objective is to serve as an interface to the real world for gathering physical information such as temperature, light, radiation, etc. Nodes perform some preliminary processing on gathered data and then send them back to a sink node for final processing. These networks are specialized in their nature, and they are decentralized and self-organized; therefore, they can be deployed without requiring the existence of a supporting infrastructure [1]. Another characteristic of these networks is that their topology is unknown a priori, so an airplane or artillery could deploy them to unreachable regions [2]. These unique features have led to wide range of applications for these networks. Healthcare monitoring, protection of critical infrastructures, monitoring environment for seismic sensing, flood and volcanic monitoring, military target tracking, and surveillance are just a few applications for these networks.

WSNs rely on wireless connectivity and usually they are unattended and without physical protections. These characteristics make them susceptible to different types of attacks. Furthermore, many applications of WSNs require secure communications. Also, the same unique features and characteristics that have promised such a broad range of applications for WSNs have rendered protocols of traditional wireless networks useless. So, new security protocols are needed for these networks. Key management as the integral part of any secure communications is not an exception and it needs even more attention. Recently, many key management schemes based on symmetric primitives have been proposed for wireless sensor networks [3, 4]. These methods have very low energy demands and they are fit for tight constraints on nodes' resources, but unfortunately they cannot provide prefect resiliency. Because WSNs are unattended and sometimes they are deployed in hostile environments, adversary could capture nodes. Also, these networks have lots of nodes. In order to keep them affordable, nodes cannot be equipped with tamper proof mechanism. All in all, adversary can collect nodes and read their information (including their cryptographic keys). So adversary can easily mount a cloning attack [5]. Therefore, a suitable key management system should be as resilient as possible.

Critical applications need perfect resiliency and PKC-based key management systems can provide it. Also, in PKC-based systems all the keys are unique and therefore detection of cloning attack would be easier. Previous studies [6, 7] have shown applicability of PKC for these networks. For example, verification of Elliptic Curve Cryptography (ECC) signature with 160-bits key takes 1.61s on ATmega128, and it consumes 45.09mj energy. More recent studies have reduced this energy consumption. For example, Imote2 which is equipped with ultralow power processor consumes only 3.51mj for the same task [8].

Following these incipient works, different PKC based systems were proposed for WSNs. First, an authentication protocol based on RSA and ECC was presented [9]. Later studies demonstrated vulnerability of this protocol to masquerade attack [10]. Later, Ren et al. proposed two different broadcast authentication methods. The first method was based on ECC and Merkle hash tree and the other one was based on Hess identity based signature [11]. Another identity-based authentication scheme reduced energy consumption of authentication [12]. Shim et al. further reduced energy consumption of identity-based authentication. This was achieved by using a pairing-optimal identity based system with message recovery method [13]. Another work employed Rabin-Williams signature for authenticating code dissemination [14].

Other researchers replaced checking authenticity of public keys with other low-cost operations. First, bloom filter and Merkle hash tree were used for this purpose [15]. Liu et al. used ECC and hash function and proposed another low-cost method for authenticating broadcast messages [16]. Recently, another method named broadcast authenticated PKC (BA) was proposed [17]. In this method, digital certificates were replaced with one-time-signatures based on delayed broadcasted messages from base station (BS). The method was three times more energy-efficient than certificate-based key management systems.

This paper presents a new set of analysis on BA protocol. First, we analyze BA protocol and show that delayed authentication of BS messages opens doors for serious denial of service attacks. To that end, we present a set of simulations and show that adversary can flood network with bogus messages. In this manner, he can exhaust memory of nodes. Also, adversary can force nodes of network to re-transmit its bogus messages. This could lead to severe energy exhaustion attack which is very serious in WSNs. Finally, we present a novel method based on hash operation and solve these problems while retaining other desirable characteristics of BA method.

The rest of this paper is organized as follows. Section II presents a brief overview of BA method and investigates its vulnerabilities to DOS attacks. Proposed method is presented in section III. Analysis of the proposed method follows in section IV. Finally the paper concludes is section V.

## II. BA METHOD AND ITS VULNERABILITY TO DOS ATTACKS

Table I describes notations used in the rest of this paper.

TABLE I. NOTATIONS USED IN THIS PAPER

| Notation | Meaning |
|---|---|
| $\|\|$ | Concatenation |
| $P_{ux}$ | Public parameter of elliptic curve Diffie-Hellman of Node $x$ |
| $P_{rx}$ | Private parameter of elliptic curve Diffie-Hellman of Node $x$ |
| $i$ | Cycle number |
| $K_{DSi}$ | Key used to generate $i$th. signature |
| $Tx_i$ | Time measured locally at node $x$ |
| $Sign_{xi}$ | $Sign_{Xi} = MAC_{K_{DSi}}(Pu_X)$ |
| $Ticket_{Xi}$ | $Ticket_{Xi} = P_{ux} \|\| Sign_{Xi}$ |
| $\Delta_i$ | Time difference between two consecutive cycles |
| $K_{Authi}$ | Key chain for authenticating BS messages |
| $MAC_K(M)$ | Message Authentication Code of message ($M$) using key ($K$) |
| $E_K(M)$ | Symmetric encryption of message ($M$) using key ($K$) |
| $K_{AB}$ | Pairwise key between node $A$ and $B$ |
| $f, g$ | Some publicly agreed on functions |
| $h(M)$ | Hash value of message $M$ |

### A. Broadcast-Authenticated (BA) method

Crudely, broadcast authenticated (BA) method relies on a set of one-time signatures generated by BS. These signatures are generated with a message authentication code (MAC), a chain of keys ({$K_{DSi}$}), and public keys ($P_{ux}$). Then, nodes are preloaded with their credentials. After nodes are deployed in the target field, they broadcast their credentials for their neighbors and wait for BS to disclose the symmetric validation key. Finally, nodes use exchanged signatures and validation key and authenticate each other. Details of BA method are as follows.

First, BS generates a key chain. This chain is used by nodes for checking authenticity of BS messages.

$$K_{Authn} \ldots \to K_{Auth1} \to K_{Auth0} \to K_{Auth00} \qquad (1)$$

Then, BS generates a second key chain. This chain is used for generating the set of one-time signatures.

$$K_{DSn} \ldots \to K_{DS1} \to K_{DS0} \qquad (2)$$

Afterwards, BS generates public ($P_{ux}$) and private ($P_{rx}$) parameters of elliptic curve Diffie-Hellman (ECDH) for every node. Then, BS generates one-time signatures of every node as:

$$Sign_{x_i} = MAC_{K_{DSi}}(P_{ux}) \quad i = 1, \ldots, n \qquad (3)$$

After nodes have exchanged their signatures, BS discloses $K_{DSi}$. Then, nodes use MAC and disclosed key for authenticating public keys of their neighbours. Finally shared key is calculated according to Diffie-Hellman scheme. Figure 1 illustrates the main steps of this protocol.

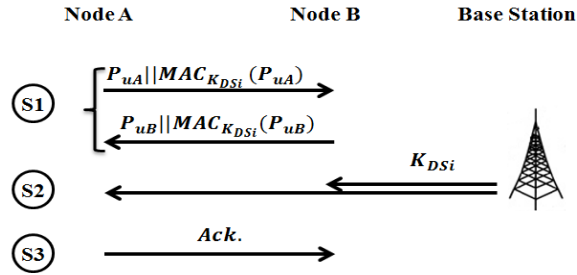

Fig. 1. Simple schematic of the BA method

Referring to figure 1, BA method consists of three main phases. First, nodes exchange their credentials with their neighbors. Second, BS broadcasts key for checking validity of those credentials ($K_{DSi}$). Finally, acknowledgment message is exchanged. Investigating the second phase of the protocol shows that it needs a mechanism for guarantying its freshness; otherwise adversary can replay old messages. The BA method relied on a symmetric primitive for this purpose. Let $i$ and $\Delta_i$ denote cycle number, and time difference between two consecutive cycles. BS uses a symmetric encryption method with $K_{Authi}$ and broadcasts the following message:

$$BS \rightarrow X: E_{K_{Authi}}(K_{DSi}||i||\Delta_i) \qquad (4)$$

After $t$ seconds, $K_{Authi}$ is broadcasted in the network.

$$BS \rightarrow X: E_{K_{Authi}} \qquad (5)$$

Now, nodes use $K_{Authi}$ and decrypt message 4. Then, they use MAC algorithm and check authenticity of public keys of their neighbor nodes. Finally, the shared key is generated by running ECDH algorithm.

### B. Vulnerabilities of BA method:

According to equation 5, authentications of message 4 is delayed for at least $t$ seconds. We show that this delay can open doors for DOS attacks. For this purpose, different scenarios were simulated.

In the BA protocol, authentications of BS messages are delayed. Adversary can exploit this characteristic and flood network with fake messages. Because nodes can only distinguish between legitimate and bogus messages after arrival of message 5, they have to buffer all received messages. This could exhaust memory of nodes and prevent them from running other tasks. To demonstrate effect of this attack we assumed that adversary chooses his time of attacks according to a uniform distribution on the interval of $[1, \tau]$. Figure 2 shows amount of exhausted memory for different values of $\tau$.

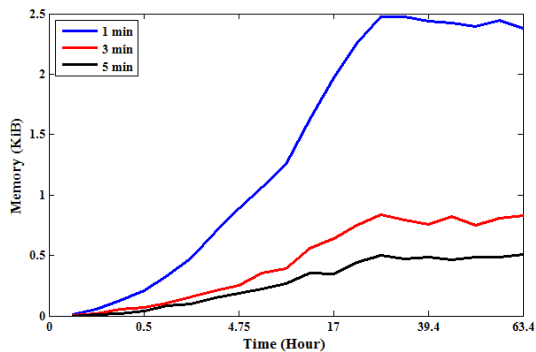

Fig. 2. Exhausted memory of a node

Apparently, as the value of $\tau$ decreases, the amount of exhausted memory increases.

Nodes of WSNs have limited transceiver range; thus, these networks rely on intermediate nodes for propagating their messages. For example, in the blind flooding, BS messages are re-transmitted by every node of network. In this fashion, messages of BS are broadcasted throughout the network. Adversary can exploit both delayed authentication and blind flooding mechanism and mount an effective energy exhaustion attack. In other words, nodes cannot authenticate bogus messages immediately. Also, due to blind flooding mechanism they will re-transmit them for their neighbors. To demonstrate effect of this attack on exhausted energy we assumed that adversary picks his time of attack according to a uniform distribution on the interval of [0, 10] minutes. Then, we calculated the total amount of energy consumed by network for re-broadcasting these fake messages. Figure 3 shows amount of exhausted energy.

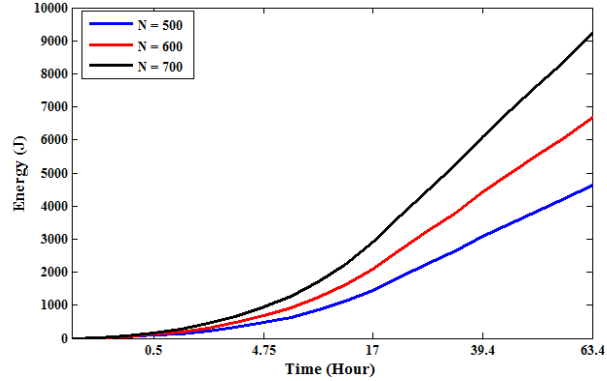

Fig. 3. Exhausted energy of network

According to figure 3 this attack is very devastating, especially for large networks.

### III. PROPOSED METHOD

Based on discussions of previous section BA method is vulnerable to DOS attacks. In this section we propose a new protocol that alleviates those problems.

Investigating equation 4 shows that it is concatenation of a key, a counter, and cycle duration. Apparently, BS knows values of counters and both keys ($K_{DSi}$, $K_{Authi}$) for all cycles. Also, BS is responsible for initiating new authentication cycles; therefore, assuming that BS (at least) knows time of next cycle is quite logical. Based on this assumption, BS has the complete knowledge about message of next cycle. Therefore, he can exploit this knowledge and append hash value of next message to the current message. In this fashion, nodes can use this auxiliary information for immediate authentication of received message. Figure 4 shows this idea.

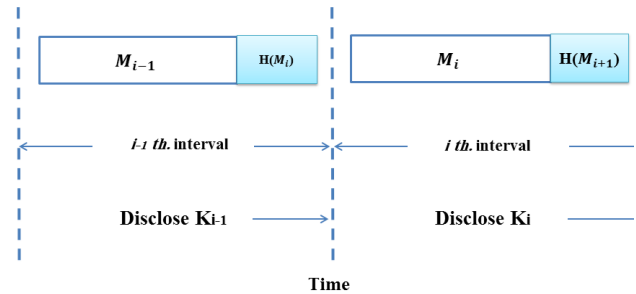

Fig. 4. Appending hash of next message to the current message

Now, we present details of the proposed method. First, instead of message 4, BS broadcasts following message.

$$BS \rightarrow X: E_{K_{Authi-1}}(K_{DSi-1}||i-1||\Delta_{i-1})||E_{K_{Authi-1}}(h(E_{K_{Authi}}(K_{DSi}||i||\Delta_i))||i-1) \qquad (6)$$

After nodes receive message 6, they split it into two parts:

$$\begin{cases} M_{i-1,1} = E_{K_{Authi-1}}(K_{DSi-1}||i-1||\Delta_{i-1}) \\ M_{i-1,2} = E_{K_{Authi-1}}(h(E_{K_{Authi}}(K_{DSi}||i||\Delta_i))||i-1) \end{cases} \qquad (7)$$

After disclosure of $K_{Authi-1}$ nodes decrypt $M_{i-1,2}$ and save value of $\mu_{i-1} = h(E_{K_{Authi}}(K_{DSi}||i||\Delta_i))$ for the next cycle. In the next cycle nodes receive message 8.

$$BS \rightarrow X: M_{i,1} || M_{i,2} \qquad (8)$$

Now, nodes immediately check authenticity of message 8 as:

$$h(M_{i,1}) \stackrel{?}{=} \mu_{i-1} \qquad (9)$$

If equation 9 holds, nodes buffer it and wait for BS to reveal $K_{Authi}$. After disclosure of this key, every node performs a hash operation and checks its authenticity. Also, freshness of message (8) is checked by comparing locally calculated time difference with the one sent from BS [19]. Then, nodes use $K_{DSi}$ with MAC operation and check validity of received public keys. Finally, nodes run ECDH and extract pairwise keys. The improved protocol with immediate authentication (*i*-BA protocol) is presented in Table II. It is noteworthy that *A, B,* and *X* denote two neighbouring nodes and every nodes of network.

TABLE II. *i*-BA PROTOCOL

| |
|---|
| $Ticket_{Xi} = [P_{uX}, Sign_{Xi}], \qquad Sign_{Xi} = MAC_{K_{DSi}}(Pu_X)$ |
| $\Delta_i = (T_{BSi} - T_{BSi-1})$ |
| $A \rightarrow B: Ticket_{Ai} \qquad B \rightarrow A: Ticket_{Bi}$ |
| $BS \rightarrow X: M_{i,1}||M_{i,2} \qquad , T_{BS} = T_{BSi}$ |
| $X: h(M_{i,1}) \stackrel{?}{=} \mu_{i-1}$ |
| $BS \rightarrow X: K_{Authi} \qquad , T_{BS} = T_{BSi} + t$ |
| $X: K_{Authi} \stackrel{h?}{\rightarrow} K_{Authi-1} \ ; T_{xi} - T_{xi-1} \stackrel{?}{\cong} \Delta_i \ ;$ |
| $A: Sign_{Bi} \stackrel{?}{=} MAC_{K_{DSi}}(Pu_B)$ |
| $B: Sign_{Ai} \stackrel{?}{=} MAC_{K_{DSi}}(Pu_A)$ |
| $A: K_{AB} = f(P_{uB} \times P_{rA}, i)$ |
| $B: K_{AB} = f(P_{uA} \times P_{rB}, i)$ |
| $A \rightarrow B : g(K_{AB})$ |

IV. ANALYSIS OF THE PROPOSED METHODS

To demonstrate potency of the proposed method different scenarios were simulated with Matlab. In these simulations nodes were randomly distributed in a 500×500 meters field. We assumed a homogeneous network with transceivers range of 30 meters. Also, the network used packet size of 41 bytes (32 bytes for payload and 9 bytes for header) [18]. Parameters of *i*-BA were as follows. 10 bits for cycle counter, 14 bits for $\Delta_i$, 128 bits for MAC and all keys, and 160 bits for ECDH keys. Finally, SHA-1 and AES methods were used for symmetric encryptions and hash operations. Each simulation was run for 100 times and then results were averaged.

*A. Security analysis*

In the *i*-BA method message 6 is the main target for adversary. We show that he cannot alter this message or replay it. Both parts of message 6 ($M_{i,1}$, $M_{i,2}$) are encrypted with a secret key and both of them are appended with a counter. If adversary manipulates message 6 these counter would decrypt into wrong counter; therefore, nodes will detect that message has been tampered.

The proposed method uses time difference between consecutive cycles. There are three possibilities considering replay of message 6.

1- Replaying message 6 in the same cycle and before disclosure of $K_{Authi}$.

2- Replaying message 6 in another cycle.

3- Replaying message 6 in the same cycle and after disclosure of $K_{Authi}$.

In the first case, adversary does not know $K_{Authi}$ so he can only re-transmit message 6 without its alteration. This is equivalent to participation of adversary in the blind flooding and network will benefit from it. In the second case, BS would disclose a different $K_{Auth}$; therefore, decrypting message from an old cycle with this new key will not pass integrity check and nodes will discard it. In the third case, adversary should jam target node so that it does not receive $K_{Authi}$ from BS. Then, he will use disclosed $K_{Authi}$ and will decrypt message 6. He may use this information and generates valid signatures for his nodes. On the other hand, he has to alter value of $\Delta_i$ in the message 6 to $\Delta_i + t$ so that target node would calculate correct time difference. But, this will contradict integrity check of equation 9. So target node will discard this message and prevent the replay attack.

*B. Connectivity of the proposed methods*

The *i*-BA method assumes that all nodes receive BS messages. Apparently for a BS with a powerful transmitter this assumption is correct; thus, the other case is investigated. In this case, nodes retransmit messages of BS for their neighbours.

Theorem1: If $k$ and $p_{loss}$ denote number of neighbours of node $C$ and probability of packet loss, probability of receiving message of BS ($p_r$) satisfies equation 10.

$$p_r = 1 - p_{loss}^{k.p_r} \qquad (10)$$

Proof: On average $k.p_r$ of neighbours of node $C$ have received message of BS. Thus, probability of node $C$ not receiving message of BS is:

$$p_{fail} = p_{loss}^{k.p_r} \qquad (11)$$

We analyzed number of neighbors in [19]. We showed that if $N$ nodes with transceiver range of $r$ are distributed uniformly over a square field of $a \times a$ then, on average nodes have $N . F_Z(\frac{r}{a})$ neighbors. Where, $F_Z$ is calculated as:

$$F_Z(z) = \begin{cases} \pi z^2 & c \in \text{Region I} \\ \pi z^2 - \frac{z^2}{2}(\theta - \sin\theta) & c \in \text{Region II} \\ \pi z^2 - \frac{z^2}{2}\left(\theta - \frac{\sin\theta}{2}\right) - \frac{z^2}{4}(\varphi - \sin\varphi) - \\ \quad \frac{z^2}{2}\tan^{-1}\left(\cot\frac{\theta}{2}\right) + z^2\cos\frac{\varphi}{2}\cos\frac{\theta}{2} & c \in \text{Region III} \end{cases} \qquad (12)$$

Also, different regions of network are depicted in figure 5.

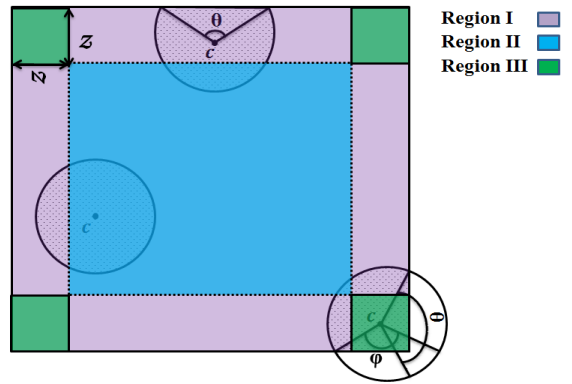

Fig. 5. Different regions of network [19]

Theorem2: If two nodes participate in $m$ authentication cycles, probability of sharing a key in the *i*-BA method is equal to:

$$P_m = 1 - (1 - p_r^4)^m \qquad (13)$$

Proof: In the *i*-BA method every node should receive two messages from BS. If nodes receive message of BS with probability of $p_r$, then probability of both nodes receiving necessary messages of BS is:

$$P_{success} = p_r^2 . p_r^2 = p_r^4 \qquad (14)$$

If nodes $A$ and $B$ participate in m authentication cycles, probability of sharing a key is:

$$P_m = 1 - (1 - P_{success})^m \qquad (15)$$

Figure 6 shows how this probability changes.

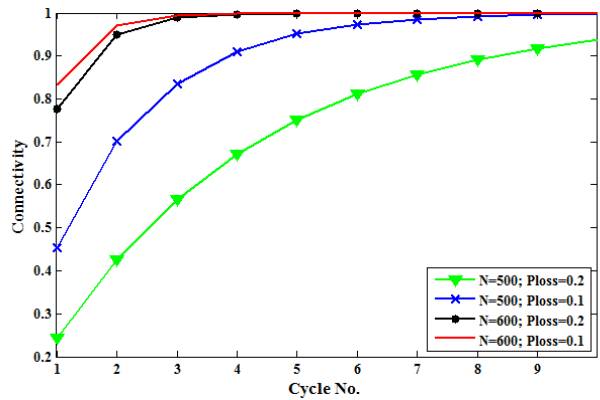

Fig. 6. Probability of sharing a key after participating in *m* authentication cycles.

According to figure 6, when packet loss is lower than 0.1 after 3 cycles, the network is fully connected.

*C. Energy consumption*

Previous studies have calculated energy consumption of transmitting data, receiving data, and executing different cryptography primitives on Mica2dot node [18, 20]. We used the same methodology of [19, 15] for calculating energy consumption. In this fashion, energy consumption for receiving necessary messages were added to energy overhead for executing different primitives. Table III presents details of computational overhead for different PKC-based key management systems.

TABLE III. COMPUTATION OVERHEAD OF DIFFERENT METHODS

| *Scheme* | Computation cost | *Reference* |
|---|---|---|
| Certificate Based | CERT verify + ECDH | [18] |
| Hybrid Method | Bloom +2 SHA1 + ECDH | [15] |
| Basic Method | 1 SHA1+1 AES+1 HMAC + ECDH | [19] |
| *i*-BA | 2 SHA1+2 AES+1 HMAC + ECDH | This paper |

After adding communication costs to the computation overhead of Table III total cost of *i*-BA becomes 60.50 mj. Table IV compares energy consumption of *i*-BA with previous works.

TABLE IV. ENERGY CONSUMPTION OF DIFFERENT PKC BASED KEY MANAGEMENT SYSTEMS

| *Scheme* | *Energy cost* | *Reference* |
|---|---|---|
| Certificate Based | 187.6 | [18] |
| Hybrid Method | 75.26 | [15] |
| Basic Method | 58.68 | [19] |
| *i*-BA | 60.50 | This paper |

In order to present a better insight into these numbers a simulation was carried out. We simulated networks with different number of nodes. Then, we calculated the amount of energy consumption for establishing a key between every node and all its neighbors. Figure 7 shows result of this simulation.

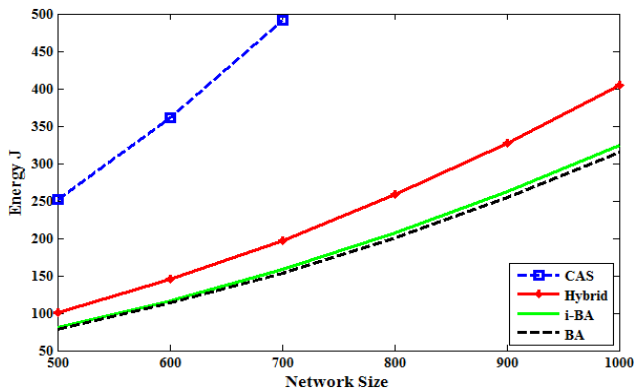

Fig.7. Total energy consumption of different PKC based key management systems.

Figure 7 shows that the proposed method is more energy efficient than previous works. Also, its energy consumption is only a little more than BA method.

*D. Key and user revocation*

If Compromised nodes and dead nodes are not handled properly, they could become serious vulnerabilities; therefore, a secure key management system should address them. The proposed method needs an auxiliary mechanism for handling compromised nodes. For, example BS could periodically broadcast ID of newly compromised nodes. On the other hand, the proposed method has an inherent mechanism for addressing dead nodes. Referring to table II, nodes uses their one-time signatures for authentication. In other words, if a nodes runs out of signatures it can no longer participate in authentication cycles. Based on this property, nodes are preloaded with adequate signatures such that when their battery depletes, there is no signature left.

*E. Scalability*

To compare scalability of different PKC based key management systems, it was assumed that nodes had 64KiB of memory on board, ECC-160 was used and node ID was 2 bytes. Certificate based and BA methods were investigated in [19] and only IDs of revoked nodes should be stored. Therefore, they can support up to 32768 nodes. In the hybrid method, BS collects all public keys of network and constructs their Merkle hash tree. Then, BS prunes this tree into a set of smaller trees and constructs a bloom filter on them. Finally, nodes are preloaded with this bloom filter. Based on [15] hybrid method can support up to 15792 nodes. At last, the proposed method (*i*-BA) has the same memory characteristics as BA. Therefore, it can support up to 32768 nodes. These results are presented in table V.

TABLE V. SCALABILITY OF DIFFERENT PKC BASED KEY MANAGEMENT SYSTEMS

| *Scheme* | Max network size | *Reference* |
|---|---|---|
| Certificate Based | 32768 | [18] |
| Hybrid Method | 15792 | [15] |
| Basic Method | 32768 | [19] |
| *i*-BA | 32768 | This paper |

V. CONCLUSION

WSNs are deployed in hostile environments and they rely on wireless communication channel; therefore, it is essential that security mechanisms be employed. Many of these security mechanisms assume that a set of key are shared between nodes of network, a task that key management system addresses. Due to tight resource constraints, many of existing key management systems are based on symmetric primitives. These methods have low energy demand but, they cannot provide perfect resiliency. Recently, an energy efficient method based on PKC was proposed. The method used a set of one time signature instead of digital certificates. In this paper we showed this method is vulnerable to DOS attacks. Then, we proposed a modified version of the protocol that achieved immediate authentication. Through simulation we showed that the improved protocol solved its vulnerability to DOS attacks while it retained other desirable characteristics of initial protocol.